\documentclass[aps,pr,twocolumn,psfig,eqsecnum,epsf]{revtex4}
\usepackage{graphicx}
\usepackage{color}

\usepackage[version=3]{mhchem} 
\usepackage{amssymb}

\begin{document}
\title{\Large \bf DC electrokinetics for spherical particles in salt-free \\ concentrated suspensions including ion size effects}
\author{ \large Rafael Roa,$^1$ F\'elix Carrique,$^{1,}$\footnote{E-mail address: carrique@uma.es} and Emilio Ruiz-Reina$^2$\\}  
\affiliation{ $^1$F\'{\i}sica Aplicada I, Universidad de M\'alaga, Spain,\\
$^2$F\'{\i}sica Aplicada II, Universidad de M\'alaga, Spain.}
\date{\today}

\vspace{0.5cm} 
\begin{abstract}                
We study the electrophoretic mobility of spherical particles and the electrical conductivity in salt-free concentrated suspensions including finite ion size effects. An ideal salt-free suspension is composed of just charged colloidal particles and the added counterions that counterbalance their surface charge. In a very recent paper [Roa \textit{et al.}, \textit{Phys. Chem. Chem. Phys.}, 2011, \textbf{13}, 3960-3968] we presented a model for the equilibrium electric double layer for this kind of suspensions considering the size of the counterions, and now we extend this work to analyze the response of the suspension under a static external electric field. The numerical results show the high importance of such corrections for moderate to high particle charges, especially when a region of closest approach of the counterions to the particle surface is considered. The present work sets the basis for further theoretical models with finite ion size corrections, concerning particularly the ac electrokinetics and rheology of such systems.
\end{abstract} 
\maketitle
\vspace{0.5cm}


\section{Introduction}

The behavior of a suspension of charged particles under a static external electric field is a subject of study of electrokinetics and constitutes a classical problem \cite{Dukhin1974,Lyklema1995,Masliyah2006}. In 1978 O'Brien and White \cite{OBrien1978} revisited the problem of electrophoresis and computed the electrophoretic mobility of a spherical particle for some cases of interest. A few years later O'Brien \cite{OBrien1981} extended their work to obtain the electrical conductivity of a dilute suspension of charged particles. Many of the classical studies concern suspensions with low particle concentration, but nowadays the concentrated regime is the one that deserves more attention because of its practical applications \cite{OBrien1990,Dukhin1999}. Ohshima developed analytic expressions for the electrophoretic mobility \cite{Ohshima1997} and the electrical conductivity \cite{Ohshima1999} in concentrated suspensions by using a cell model approach. This approach has been successfully tested against experimental electrokinetic results in concentrated suspensions \cite{Carrique2003b,Delgado2005,Cuquejo2006,Reiber2007}. From a theoretical point of view, these systems are difficult to understand due to the inherent complexity associated with the increasing particle-particle electrohydrodynamic interactions as particle concentration grows, and the possibility of overlapping between adjacent double layers which will be unavoidably present with high particle concentrations \cite{Carrique2003}.

In many typical cases, the presence of an external salt added to the system gives rise to an effective screening effect on repulsive electrostatic particle-particle interactions, depending on the salt concentration, which are mainly responsible, for example, for the generation of colloidal crystals or glasses. Thus, it would be of worth to study systems with a low screening regime for such interactions. Those systems are named salt-free because of the absence of added external salt. The formation of colloidal crystals is easier in this kind of systems, even at sufficiently low particle volume fractions \cite{Sood1991,Medebach2003,Palberg2004,Medebach2004}. Of course, these salt-free systems contain ions in solution, the so-called ``added counterions'' stemming from the particles as they get charged, that counterbalance their surface charge preserving the electroneutrality. With the help of a cell model approach, Ohshima \cite{Ohshima2002}, and later Chiang \textit{et al.} \cite{Chiang2006} and Carrique \textit{et al.} \cite{Carrique2006} stressed the study of the electrophoretic mobility of spherical particles in salt-free suspensions. In the case of Carrique \textit{et al.}, the study was also extended to the computation of the electrical conductivity in salt-free concentrated suspensions. On the other hand, there is a lack of experimental results concerning the electrokinetic properties of salt-free systems due to its difficult preparation. Between them, the experimental work of Palberg and coworkers \cite{Medebach2003,Palberg2004,Medebach2004} is probably the most extensive using this kind of systems. 

All these theoretical studies are based on a mean-field description, that has a reasonable success when representing the ionic concentration profiles at low to moderately charged interfaces. These studies also consider point-like ions, which, for highly charged particles, yields unphysical high counterion concentration profiles near such interfaces. In addition, these treatments neglect ion-ion correlations, which simplifies the real scenario. We can find in the literature different attempts to overcome these limitations. Some of them concern microscopic descriptions of ion-ion correlations and the finite size of the ions \cite{LozadaCassou1999,Lobaskin2007,Chatterji2007,MartinMolina2009,Pagonabarraga2010} that are able to predict important phenomena, like overcharging \cite{Lyklema2009}, but are basically restricted to equilibrium conditions. Others are based on macroscopic descriptions considering average interactions by mean-field approximations that include entropic contributions related to the excluded volume effect when the ions have a finite size \cite{Bikerman1942,KraljIglic1996,Borukhov1997,Borukhov2004,Kilic2007,ArandaRascon2009,LopezGarcia2010}. The interested reader can find a historical overview about steric effects in the review of Bazant \textit{et al.} \cite{Bazant2009}. Some simulations results showed that these corrections work appreciably well with monovalent electrolytes for high surface charge densities and/or large ionic sizes \cite{IbarraArmenta2009}. On the other hand, the macroscopic approaches permit us to make predictions under equilibrium and non-equilibrium conditions. Some authors \cite{ArandaRascon2009b,Khair2009,LopezGarcia2011} have extended their works for equilibrium conditions to predict non-equilibrium properties, like the electrophoretic mobility or the electrical conductivity in diluted suspensions with electrolytes. 

One of the classical drawbacks of the mean-field approaches without ion-ion correlations is their inability to explain important phenomena like overcharging, while Monte Carlo simulations achieve it by considering full ion-ion correlations. Very recently, L\'opez-Garc\'ia \textit{et al.} \cite{LopezGarcia2010,LopezGarcia2011} presented a modified standard electrokinetic model for diluted suspensions which takes into account the finite ion size and considers a minimum approach distance of ions to the particle surface  not necessarily equal to their effective radius in the bulk solution. They show that this model can predict overcharging in the case of high electrolyte concentrations and counterion valence. We think that this is a very important result because to our knowledge this is the first time that a phenomenological theory based on macroscopic descriptions is able to predict this phenomenon.

In a very recent paper \cite{Roa2011} we presented a model for the equilibrium electric double layer for spherical particles in salt-free concentrated suspensions considering the size of the counterions. The procedure was a generalization of that already used by Borukhov \cite{Borukhov2004} for the special case of a salt-free suspension valid for the concentrated case. Unlike Borukhov's treatment, our model also incorporates an excluded region in contact with the particle of a hydrated radius size, which has been shown by Aranda-Rasc\'on \textit{et al.} \cite{ArandaRascon2009,ArandaRascon2009b} to yield a more realistic representation of the solid-liquid interface, and also to predict results in better agreement with experimental electrokinetic data. 

Our aim in this paper is to extend our previous work valid for equilibrium conditions to analyze the response of a salt-free concentrated suspension under a static external electric field considering the size of the counterions. We will study specially the electrophoretic mobility of the particles and the electrical conductivity of the suspension. We will follow the treatment developed by Carrique \textit{et al.} \cite{Carrique2006} for salt-free concentrated suspensions with point-like ions to achieve our electrokinetic model with ion size effects.

The plan of this paper is as follows. In Section \ref{sec:ekeq} we modify the governing electrokinetic equations to include the size of the counterions and their distance of closest approach to the particle surface. The boundary conditions needed to solve the problem are discussed in Section \ref{sec:bc}. In Sections \ref{sec:mob} and \ref{sec:cond} we present the expressions for the calculation of the electrophoretic mobility and the electrical conductivity, respectively. The results of the numerical calculations are shown in Section \ref{sec:results} and analyzed upon changing particle volume fraction, particle surface charge density, and size of the counterions. In order to show the realm of the finite ion size effect in salt-free suspensions, the results will be compared with the predictions that do not take into account a finite distance of closest approach to the particle surface, and also with the standard predictions for point-like ions. Conclusions are presented in Section \ref{sec:conclusions}.

\section{Theory} \label{theory}

\subsection{Electrokinetic equations} \label{sec:ekeq}
\begin{figure}[t]
\centering
  \includegraphics[height=4.5cm]{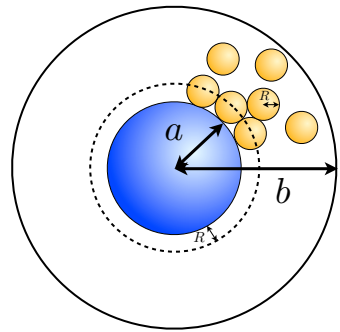}
  \caption{Cell model including the distance of closest approach of the counterions to the particle surface.}
  \label{fgr:cell}
\end{figure}

We use a cell model approach to account for the interactions between particles in concentrated suspensions through adequately chosen boundary conditions (bare Coulomb interactions among particles are included in an average sense, but ions-induced interactions between particles as well as ion-ion correlations, are ignored). For details about the cell model approach see the excellent review of Zholkovskij \textit{et al.} \cite{Zholkovskij2007}. In this approach, represented in Fig. \ref{fgr:cell}, each spherical particle of radius $a$ is surrounded by a concentric shell of the liquid medium, having an outer radius $b$ such that the particle/cell volume ratio in the cell is equal to the particle volume fraction throughout the entire suspension, that is \cite{Happel1958}
\begin{equation}\label{phidef}
\phi=\left( \frac{a}{b}\right)^3
\end{equation}

The basic assumption of the cell model is that the macroscopic properties of a suspension can be obtained  from appropriate averages of local properties in a unique cell.

Let us consider a spherical charged particle of radius $a$, surface charge density $\sigma$ and relative permittivity $\epsilon_{rp}$ immersed in a salt-free medium of relative permittivity $\epsilon_{rs}$ and viscosity $\eta$, with only the presence of the added counterions of valence $z_c$ and drag coefficient $\lambda_c$. In the presence of a static electric field $\mathbf{E}$ the particle moves with a uniform velocity $\mathbf{v}_e$, the electrophoretic velocity. The axes of the spherical coordinate system ($r$, $\theta$, $\varphi$) are fixed at the center of the particle, with the polar axis ($\theta=0$) parallel to the electric field. The solution of the problem requires the knowledge, at every point $\mathbf{r}$ of the system, of the electric potential, $\Psi(\mathbf{r})$, the number density of counterions, $n_c(\mathbf{r})$, their drift velocity, $\mathbf{v}_c(\mathbf{r})$, and the pressure, $P(\mathbf{r})$. The fundamental electrokinetic equations connecting them are \cite{OBrien1978}: the Poisson equation for the relationship between the electric potential and the charge density, 
\begin{equation}\label{poisson}
  \nabla^2\Psi(\mathbf{r})=-\frac{z_ce}{\epsilon_0\epsilon_{rs}}n_c(\mathbf{r})
\end{equation}
the Navier-Stokes equation for low Reynolds number in the presence of an electrical body force for the fluid velocity,
\begin{equation}
  \eta\nabla^2\mathbf{v}(\mathbf{r})-\nabla P(\mathbf{r})-z_cen_c(\mathbf{r})\nabla \Psi(\mathbf{r})=0
\end{equation}
the continuity equation for the counterions that implies the conservation of the number of counterions in the system,
\begin{equation}
  \nabla\cdot[n_c(\mathbf{r}) \mathbf{v}_c(\mathbf{r})]=0
\end{equation}
and the Nernst-Planck equation for the flow of the counterions,
\begin{equation}\label{nersnt-p}
  n_c(\mathbf{r})\mathbf{v}_c(\mathbf{r})=n_c(\mathbf{r})\mathbf{v}(\mathbf{r})-\frac{1}{\lambda_c}n_c(\mathbf{r})\nabla\mu_c(\mathbf{r})
\end{equation}
where $\mu_c(\mathbf{r})$ is the electrochemical potential of the counterions. We also take into account the continuity equation for an incompressible fluid flow,
\begin{equation}\label{incompress}
  \nabla\cdot\mathbf{v}(\mathbf{r})=0
\end{equation}

In these equations, $\epsilon_0$ is the vacuum permittivity and $e$ is the elementary electric charge. The drag coefficient $\lambda_c$ is related to the limiting ionic conductance $\Lambda_c^0$ or the diffusion coefficient $D_c$ by
\begin{equation}
  \lambda_c=\frac{N_Ae^2|z_c|}{\Lambda_c^0}=\frac{k_BT}{D_c}
\end{equation}
where $N_A$ is Avogadro's number, $k_B$ is Boltzmann's constant, and $T$ is the absolute temperature.

As we are interested in studying the linear response of the system to an electric field, the following perturbation scheme is applied, where each quantity $X$ is written as the sum of its equilibrium value, $X^0$, plus a perturbation term $\delta X$ linearly dependent with the field:
\begin{align}
  & \Psi(\mathbf{r})=\Psi^0(r)+\delta\Psi(\mathbf{r}) \nonumber \\
  & n_c(\mathbf{r})=n_c^0(r)+\delta n_c(\mathbf{r}) \nonumber \\
  & \mu_c(\mathbf{r})=\mu_c^0+\delta \mu_c(\mathbf{r}) 
\end{align}

As we showed in a previous paper \cite{Roa2011}, we introduce the finite size of the counterions by considering their excluded volume and including the entropy of the solvent molecules in the free energy of the suspension, $F=U-TS$
\begin{multline}
  U=\int \mathrm{d}\textbf{r}\bigg{[}-\frac{\epsilon_0\epsilon_{rs}}{2}|\nabla\Psi^0(\mathbf{r})|^2\\ +z_cen_c^0(\mathbf{r})\Psi^0(\mathbf{r})-\mu_c^0n_c^0(\mathbf{r})\bigg{]}
\end{multline}
\begin{multline}\label{entropy}
  -TS=k_BTn_c^{max}\int \mathrm{d}\textbf{r}\bigg{[}\frac{n_c^0(\mathbf{r})}{n_c^{max}}\ln \left(\frac{n_c^0(\mathbf{r})}{n_c^{max}}\right)\\ 
  +\left(1-\frac{n_c^0(\mathbf{r})}{n_c^{max}}\right)\ln \left(1-\frac{n_c^0(\mathbf{r})}{n_c^{max}}\right)\bigg{]}
\end{multline}
being $n_c^{max}$ the maximum possible concentration of counterions due to the excluded volume effect, defined as $n_c^{max}=V^{-1}$, where $V$ is the average volume occupied by an ion in the solution. The last term in eqn (\ref{entropy}) is the one that accounts for the ion size effect, and was proposed earlier by Borukhov et al. \cite{Borukhov1997}.

The variation of the free energy $F=U-TS$ with respect to $\Psi^0(\mathbf{r})$ provides the Poisson equation for the equilibrium
\begin{equation}\label{poisson0}
  \nabla^2\Psi^0(\mathbf{r})=-\frac{z_ce}{\epsilon_0\epsilon_{rs}}n_c^0(\mathbf{r})
\end{equation}
and the equilibrium counterions concentration is obtained performing the variation of the free energy with respect to $n_c^0(\mathbf{r})$, obtaining
\begin{equation}\label{nc0}
  n_c^0(\mathbf{r})=\frac{b_c\exp\left(-\frac{z_ce\Psi^0(\mathbf{r})}{k_BT}\right)}{1+\frac{b_c}{n_c^{max}}\left[\exp\left(-\frac{z_ce\Psi^0(\mathbf{r})}{k_BT}\right)-1\right]}
\end{equation}
where $b_c$ is an unknown coefficient that represents the ionic concentration where the electric potential is chosen to be zero.

We also obtain the electrochemical potential doing the variation of the free energy with respect to $n_c^0(\mathbf{r})$ \cite{Kilic2007}, and assuming that the electrochemical potential out of equilibrium can be expressed in a similar way that in equilibrium conditions
\begin{equation}\label{echempot}
  \mu_c(\mathbf{r})=z_ce\Psi(\mathbf{r})+k_BT\ln \left(\frac{\frac{n_c(\mathbf{r})}{n_c^{max}}}{1-\frac{n_c(\mathbf{r})}{n_c^{max}}}\right)
\end{equation}

In the case of equilibrium there is no external field and the particle is surrounded by a spherically symmetrical charge distribution. Applying this symmetry and combining eqn (\ref{poisson0}) and (\ref{nc0}), we obtain a modified Poisson-Boltzmann equation for the equilibrium electric potential
\begin{multline}\label{mpb0}
\frac{\mathrm{d}^2\Psi^0(r)}{\mathrm{d}r^2}+\frac{2}{r}\frac{\mathrm{d}\Psi^0(r)}{\mathrm{d}r}\\ =-\frac{z_ce}{\epsilon_0\epsilon_{rs}}\frac{b_c\exp\left(-\frac{z_ce\Psi^0(r)}{k_BT}\right)}{1+\frac{b_c}{n_c^{max}}\left[\exp\left(-\frac{z_ce\Psi^0(r)}{k_BT}\right)-1\right]}
\end{multline}

The electroneutrality of the cell implies that
\begin{equation}\label{electroneu}
Q=4\pi a^2\sigma=-4\pi z_ce \int_a^b n_c^0 (r)r^2\mathrm{d}r 
\end{equation}
which is a necessary expression for the iterative calculation of the unknown $b_c$ coefficient. An interested reader can find more details about the modified Poisson-Boltzmann equation for the equilibrium in ref. \cite{Roa2011}.

As indicated before, it is convenient to write the non-equilibrium quantities in terms of their equilibrium values plus a field-dependent perturbation. The symmetry of the problem allows us to define the functions $h(r)$, $\phi_c(r)$, and $Y(r)$ \cite{Ohshima1997}
\begin{multline}\label{vOhshima}
  \mathbf{v}(\mathbf{r})=(v_r,v_\theta,v_\phi)=\\ \left(-\frac{2}{r}hEcos\theta,\frac{1}{r}\frac{\mathrm{d}}{\mathrm{d}r}(rh)E\sin\theta,0\right)
\end{multline}
\begin{equation}\label{deltamuc}
  \delta\mu_c(\mathbf{r})=-z_ce\phi_c(r)E\cos\theta
\end{equation}
\begin{equation}\label{deltapsi}
  \delta\Psi(\mathbf{r})=-Y(r)E\cos\theta
\end{equation}
with $E=|\mathbf{E}|$.

Substituting into the differential electrokinetic equations, eqn (\ref{poisson})-(\ref{incompress}), the above mentioned perturbation scheme, neglecting nonlinear perturbations terms, and making use of the symmetry conditions of the problem we obtain
\begin{multline}\label{L_navier-stokes}
  \mathcal{L}(\mathcal{L}h(r))=-\frac{z_ce^2}{k_BT\eta r}\left(\frac{\mathrm{d}\Psi^0(r)}{\mathrm{d}r}\right)\\ \times n_c^0(r)\left(\phi_c(r)-\frac{n_c^0(r)}{n_c^{max}}Y(r)\right)
\end{multline}
\begin{multline}\label{L_continuity}
  \mathcal{L}\phi_c(r)=\frac{e}{k_BT}\left(\frac{\mathrm{d}\Psi^0(r)}{\mathrm{d}r}\right)\\ \times \left(1-\frac{n_c^0(r)}{n_c^{max}}\right)\left(z_c\frac{\mathrm{d}\phi_c(r)}{\mathrm{d}r}-\frac{2\lambda_c}{e}\frac{h(r)}{r}\right)
\end{multline}
\begin{equation}\label{L_poisson}
  \mathcal{L}Y(r)=-\frac{z_c^2e^2n_c^0(r)}{\epsilon_0\epsilon_{rs}k_BT}\left(\phi_c(r)-Y(r)\right)
\end{equation}
where the $\mathcal{L}$ operator is defined by
\begin{equation}
  \mathcal{L} \equiv \frac{\mathrm{d}^2}{\mathrm{d}r^2}+\frac{2}{r}\frac{\mathrm{d}}{\mathrm{d}r}-\frac{2}{r^2}
\end{equation}

If we consider point-like counterions, $n_c^{max}=\infty$, eqn (\ref{mpb0}), (\ref{L_navier-stokes})-(\ref{L_poisson}) become the expressions obtained by Carrique \textit{et al.} \cite{Carrique2006}.

Following the work of Aranda-Rasc\'on \textit{et al.} \cite{ArandaRascon2009}, we incorporate a distance of closest approach of the counterions to the particle surface, resulting from their finite size. We assume that counterions cannot come closer to the surface of the particle than their effective hydration radius, $R$, and, therefore, the ionic concentration will be zero in the region between the particle surface, $r=a$, and the spherical surface, $r=a+R$, defined by the counterion effective radius. This reasoning implies that counterions are considered as spheres of radius $R$ with a point charge at its center.

With this consideration, the electrokinetic equations needed to solve the problem change into the following stepwise equations: the Poisson equation for the equilibrium electric potential becomes
\begin{equation}\label{mpb0_LAP}
\begin{cases}
\frac{\mathrm{d}^2\Psi^0(r)}{\mathrm{d}r^2}+\frac{2}{r}\frac{\mathrm{d}\Psi^0(r)}{\mathrm{d}r}=0 & \text{ \ } a\leq r \leq a+R \\
& \\
\text{Eqn }(\ref{mpb0}) & \text{ \ } a+R\leq r \leq b
\end{cases}
\end{equation}
the Navier-Stokes equation for the fluid velocity turns into
\begin{equation}\label{L_navier-stokes_LAP}
\begin{cases}
\mathcal{L}(\mathcal{L}h(r))=0 & \text{ \ } a\leq r \leq a+R \\
& \\
\text{Eqn }(\ref{L_navier-stokes}) & \text{ \ } a+R\leq r \leq b
\end{cases}
\end{equation}
the equation for the conservation of the number of counterions now reads
\begin{equation}\label{L_continuity_LAP}
\begin{cases}
\phi_c(r)=0 & \text{ \ } a\leq r \leq a+R \\
& \\
\text{Eqn }(\ref{L_continuity}) & \text{ \ } a+R\leq r \leq b
\end{cases}
\end{equation}
and the Poisson equation for the perturbation of the electric potential changes into
\begin{equation}\label{L_poisson_LAP}
\begin{cases}
\mathcal{L}Y(r)=0 & \text{ \ } a\leq r \leq a+R \\
& \\
\text{Eqn }(\ref{L_poisson}) & \text{ \ } a+R\leq r \leq b
\end{cases}
\end{equation}

\subsection{Boundary conditions} \label{sec:bc}
We next specify the boundary conditions we use for the resolution of the electrokinetic equations. In the case of the equilibrium electric potential, we fix its origin at $r=b$, which results in 
\begin{equation}
\Psi^0(b)=0
\end{equation}

Using the electroneutrality condition of the cell, eqn (\ref{electroneu}), and Gauss theorem to the outer surface of the cell, we obtain
\begin{equation}
\frac{\mathrm{d}\Psi^0(r)}{\mathrm{d}r}\bigg |_{r=b}=0
\end{equation}

On the other hand, specifying the electrical state of the particle, and applying Gauss theorem to the outer side of the particle surface $r=a$ we get
\begin{equation}
\frac{\mathrm{d}\Psi^0(r)}{\mathrm{d}r}\bigg |_{r=a}=-\frac{\sigma}{\epsilon_0\epsilon_{rs}}
\end{equation}

We also force the equilibrium potential and its first derivative to be continuous at the surface $r=a+R$ defined by the counterion effective radius.

In the case of the electric potential out of equilibrium, the discontinuity of the normal component of the displacement vector at the particle surface of charge density $\sigma$ states
\begin{equation}\label{dispvec}
  \epsilon_{rs}\nabla\Psi(\mathbf{r})\cdot \mathbf{\hat{r}}\big |_{r=a}-\epsilon_{rp}\nabla\Psi_P(\mathbf{r})\cdot \mathbf{\hat{r}}\big |_{r=a}=\frac{-\sigma}{\epsilon_0}
\end{equation}
where $\Psi_P(\mathbf{r})$ is the electric potential in the interior region of the solid particle, and $\mathbf{\hat{r}}$ is the normal unit vector outward to the surface. Also, the continuity of the electric potential at the surface of the particle has to be considered
\begin{equation}
  \Psi(\mathbf{r})\big |_{r=a}=\Psi_P(\mathbf{r})\big |_{r=a}
\end{equation}

According to Shilov-Zharkikh-Borkovskaya boundary conditions \cite{Shilov1981}, the connection between the macroscopic experimentally measured electric field $\langle\mathbf{E}\rangle$ and local electric properties is
\begin{equation}
  \Psi(\mathbf{r})\big |_{r=b}-\Psi^0(r)\big |_{r=b}=-\langle\mathbf{E}\rangle\cdot\mathbf{r}\big |_{r=b}
\end{equation}

We also must impose the continuity of the electric potential out of equilibrium and of its first derivative at the boundary surface $r=a+R$.

Following again Shilov-Zharkikh-Borkovskaya boundary conditions, the ionic perturbation at the outer surface of the cell must be zero, or equivalently
\begin{equation}\label{shilovzarbor}
  n_c(\mathbf{r})\big |_{r=b}=n_c^0(r)\big |_{r=b}
\end{equation}

As the solid particles are impenetrable objects for the ions, the velocity of the ions in the normal direction to the particle surface is zero
\begin{equation}
  \mathbf{v}_c(\mathbf{r})\cdot\mathbf{\hat{r}}\big |_{r=a+R}=0
\end{equation}

Due to the inclusion of a distance of closest approach of the counterions to the particle surface, the density of counterions $n_c(\mathbf{r})$ and their drift velocity $\mathbf{v}_c(\mathbf{r})$ will be discontinuous at the surface $r=a+R$, being zero in the region $[a,a+R]$ and non-zero in the region $[a+R,b]$.

The liquid located at the particle surface, $r=a$, is considered immobile, strongly attached to the particle. The latter fact that the liquid cannot slip on the particle is expressed as
\begin{equation}
  \mathbf{v}(\mathbf{r})\big |_{r=a}=0
\end{equation}

At the outer surface of the cell, $r=b$, we follow Kuwabara's boundary conditions \cite{Kuwabara1959}. In the radial direction, the velocity of the liquid far from the particle will be the negative of the radial component of the electrophoretic velocity
\begin{equation}\label{bcmob}
  \mathbf{v}(\mathbf{r})\cdot\mathbf{\hat{r}}\big |_{r=b}=-\mathbf{v}_e(\mathbf{r})\cdot\mathbf{\hat{r}}\big |_{r=b}
\end{equation}

According also to Kuwabara, the fluid flow is free of vorticity at the outer surface of the cell
\begin{equation}\label{vorticity}
  \nabla\times\mathbf{v}(\mathbf{r})\big |_{r=b}=0
\end{equation}

At the boundary surface $r=a+R$ we must consider the continuity of the normal and tangential components of the fluid velocity as well as the continuity of vorticity and pressure \cite{Ahualli2009}.

Finally, in the stationary state, the net force acting on the particle or the unit cell must be zero. Since the net electric charge within the unit cell is zero, there is no net electric force acting on the unit cell, and we need to consider only the hydrodynamic force. For details about this boundary condition see ref. \cite{Ohshima1997} or Appendix 1 in ref. \cite{Carrique2008}.

In terms of the radial functions $\Psi^0(r)$, $Y(r)$, $\phi_c(r)$ and $h(r)$, the previous boundary conditions change into:

\noindent
(\textit{i}) at the particle surface $r=a$
\begin{equation}
\frac{\mathrm{d}\Psi^0(r)}{\mathrm{d}r}\bigg |_{r=a}=-\frac{\sigma}{\epsilon_0\epsilon_{rs}}
\end{equation}
\begin{equation}
  \frac{\mathrm{d}Y(r)}{\mathrm{d}r}\bigg |_{r=a}-\frac{\epsilon_{rp}}{\epsilon_{rs}}\frac{Y(a)}{a}=0
\end{equation}
\begin{equation}
h(a)=0
\end{equation}
\begin{equation}
\frac{\mathrm{d}h(r)}{\mathrm{d}r}\bigg |_{r=a}=0
\end{equation}
(\textit{ii}) at the surface $r=a+R$ defined by the counterion effective radius
\begin{equation}
\Psi^0(a+R^-)=\Psi^0(a+R^+)
\end{equation}
\begin{equation}
\frac{\mathrm{d}\Psi^0(r)}{\mathrm{d}r}\bigg |_{r=a+R^-}=\frac{\mathrm{d}\Psi^0(r)}{\mathrm{d}r}\bigg |_{r=a+R^+}
\end{equation}
\begin{equation}
Y(a+R^-)=Y(a+R^+)
\end{equation}
\begin{equation}
\frac{\mathrm{d}Y(r)}{\mathrm{d}r}\bigg |_{r=a+R^-}=\frac{\mathrm{d}Y(r)}{\mathrm{d}r}\bigg |_{r=a+R^+}
\end{equation}
\begin{equation}
\frac{\mathrm{d}\phi_c(r)}{\mathrm{d}r}\bigg |_{r=a+R^+}=0
\end{equation}
\begin{equation}
h(a+R^-)=h(a+R^+)
\end{equation}
\begin{equation}
\frac{\mathrm{d}h(r)}{\mathrm{d}r}\bigg |_{r=a+R^-}=\frac{\mathrm{d}h(r)}{\mathrm{d}r}\bigg |_{r=a+R^+}
\end{equation}
\begin{equation}
\mathcal{L}h(a+R^-)=\mathcal{L}h(a+R^+)
\end{equation}
\begin{multline}
  \frac{\mathrm{d^3}h(r)}{\mathrm{d}r^3}\bigg |_{r=a+R^-}=\frac{\mathrm{d^3}h(r)}{\mathrm{d}r^3}\bigg |_{r=a+R^+}\\  -\frac{z_ce}{\eta(a+R)}n_c^0(a+R^+)Y(a+R^+)
\end{multline}
(\textit{iii}) and finally, at the outer surface of the cell $r=b$
\begin{equation}
\Psi^0(b)=0
\end{equation}
\begin{equation}
\frac{\mathrm{d}\Psi^0(r)}{\mathrm{d}r}\bigg |_{r=b}=0
\end{equation}
\begin{equation}
Y(b)=b
\end{equation}
\begin{equation}
\phi_c(b)=b
\end{equation}
\begin{equation}
\mathcal{L}h(b)=0
\end{equation}
\begin{equation}
\eta\frac{\mathrm{d}}{\mathrm{d}r}\big[r\mathcal{L}h(r)\big]_{r=b}-z_ceb_cY(b)=0
\end{equation}

\subsection{Electrophoretic mobility} \label {sec:mob}
The electrophoretic mobility $\mu$ of a spherical particle in a concentrated colloidal suspension can be defined from the relation between the electrophoretic velocity of the particle $\mathbf{v}_e$ and the macroscopic electric field $\langle\mathbf{E}\rangle$. From the boundary condition, eqn (\ref{bcmob}), the definition $|\mathbf{v}_e|=\mu|\langle\mathbf{E}\rangle|$, and the symmetry eqn (\ref{vOhshima}), we obtain
\begin{equation}
\mu=\frac{2h(b)}{b}
\end{equation}

As usual, the mobility data will be scaled as 
\begin{equation}
\mu^*=\frac{3\eta e}{2\epsilon_0\epsilon_{rs}k_BT}\mu
\end{equation}
where $\mu^*$ is the nondimensional electrophoretic mobility.

\subsection{Electrical conductivity} \label {sec:cond}
The electrical conductivity, $K$, of the suspension, is usually defined in terms of the volume averages of the local electric current density and electric field in a cell representing the whole suspension.
\begin{equation}
\langle\mathbf{J}\rangle=\frac{1}{V_{cell}}\int_{V_{cell}}\mathbf{J}(\mathbf{r})\mathrm{d}V=K\langle\mathbf{E}\rangle
\end{equation}

The macroscopic electric field $\langle\mathbf{E}\rangle$ is given by
\begin{equation}
\langle\mathbf{E}\rangle=-\frac{1}{V_{cell}}\int_{V_{cell}}\nabla\Psi(\mathbf{r})\mathrm{d}V
\end{equation}

The electric current density of a salt-free suspension is defined by
\begin{multline}
\mathbf{J}(\mathbf{r})=z_cen_c(\mathbf{r})\mathbf{v}_c(\mathbf{r})\\ =z_cen_c\left(\mathbf{v}(\mathbf{r})-\frac{1}{\lambda_c}\nabla\mu_c(\mathbf{r})\right)
\end{multline}

Following a similar procedure to that described for the conductivity of suspensions in salt solutions in ref. \cite{Carrique2005} we obtain (see also ref. \cite{Carrique2006})
\begin{equation}
K=\left(\frac{z_c^2e^2}{\lambda_c}\frac{\mathrm{d}\phi_c(r)}{\mathrm{d}r}\bigg |_{r=b}-\frac{2h(b)}{b}z_ce\right)\frac{b}{Y(b)}n_c^0(b)
\end{equation}

\section{Results and discussion} \label{sec:results}

We will discuss the results obtained from three different electrokinetic models: the classical with point-like counterions (PL) from ref. \cite{Carrique2006}, its modification to take into account the finite ion size (FIS), eqn (\ref{mpb0}), (\ref{L_navier-stokes})-(\ref{L_poisson}), and the complete electrokinetic model that also considers the distance of closest approach of the counterions to the charged particle surface (FIS+L), eqn (\ref{mpb0_LAP})-(\ref{L_poisson_LAP}). In this complete electrokinetic model the electric potential fulfills the Laplace equation in the excluded region in contact with the particle.

The governing electrokinetic equations with their boundary conditions form a boundary value problem that can be solved numerically using the MATLAB routine bvp4c \cite{Kierzenka2001}. This routine computes the solution with a finite difference method by the three-stage Lobatto IIIA formula, which is a collocation method that provides a $C^1$ solution that is fourth order uniformly accurate at all the mesh points. The resulting mesh is non-uniformly spaced out and has been chosen to fulfill the admitted error tolerance (always taken lower than $10^{-5}$).

For all the calculations, the temperature $T$ has been chosen equal to 298.15 K, the viscosity of the solution $\eta=0$.89$\cdot10^{-3}$ Pa$\cdot$s, and the relative electric permittivity of the suspending liquid $\epsilon_{rs}=$ 78.55, which coincides with that of the deionized water, although no additional ions different to those stemming from the particles have been considered in the present model. We have used the value  $\epsilon_{rp}=2$ for the relative permittivity of the particles. Also, the particle radius $a$ has been taken equal to 100 nm and the valence of the added counterions $z_c=+1$. Other values for $z_c$ could have been chosen. The model for point-like ions is able to work with any value of $z_c$, but we think that the predictions of this model will be less accurate in the case of multivalent counterions, since it is based on a mean-field approach that does not consider ion-ion correlations. Nevertheless, when we take into account the finite size of the ions, the main objective of this work, we include correlations associated with the ionic excluded volume, solving partly this problem, because we are still not considering the electrostatic ion-ion correlations. 

For the sake of simplicity, we assume that  the average volume occupied by a counterion is $V=(2R)^{3}$, being $2R$ the counterion effective diameter. With this consideration, the maximum possible concentration of counterions due to the excluded volume effect is $n_c^{max}=(2R)^{-3}$. This corresponds to a simple cubic package (52\% packing). In molar concentrations, the values used in the calculations, $n_c^{max}=$ 22, 4 and 1.7 M, correspond approximately to counterion effective diameters of $2R=$ 0.425, 0.75 and 1 nm, respectively. These are typical hydrated ionic radii \cite{Israelachvili1992}. The diffusion coefficient of the counterions has been chosen as $D_c=9$.34$\cdot10^{-9}$ m$^2$/s, that corresponds to the value for H$^+$ ions, which are commonly found in many experimental conditions with pure salt-free suspensions, although other different values could have been used.

\subsection{Electrophoretic mobility}

\begin{figure}[t]
\centering
  \includegraphics[width=8.3cm]{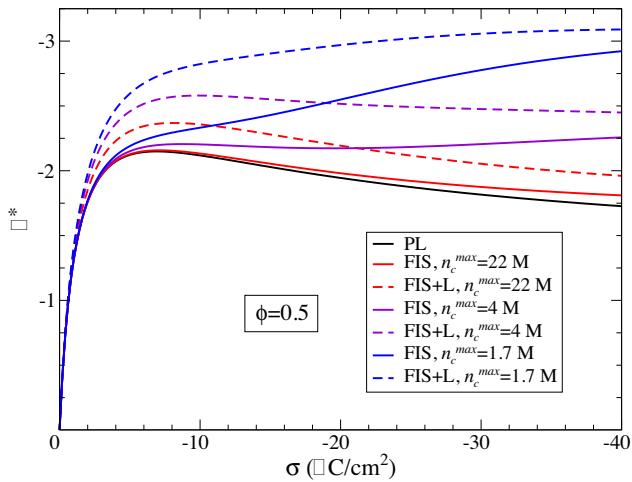}
  \caption{Scaled electrophoretic mobility against the particle surface charge density for different ion sizes, considering (dashed lines) or not (solid lines) the excluded region in contact with the particle. Black lines show the results for point-like ions.}
  \label{fgr:mob_sigmaphi05}
\end{figure}

\begin{figure}[t]
\centering
  \includegraphics[width=8.3cm]{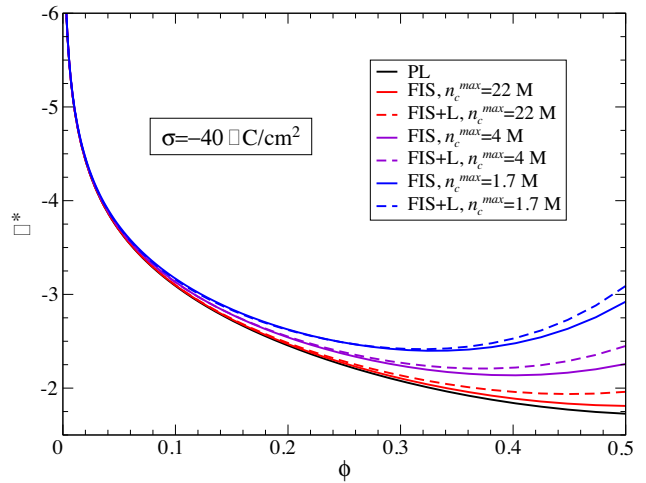}
  \caption{Scaled electrophoretic mobility against the particle volume fraction for different ion sizes, considering (dashed lines) or not (solid lines) the excluded region in contact with the particle. Black lines show the results for point-like ions.}
  \label{fgr:mob_phisigma40}
\end{figure}

\begin{figure}[t]
\centering
  \includegraphics[width=8.3cm]{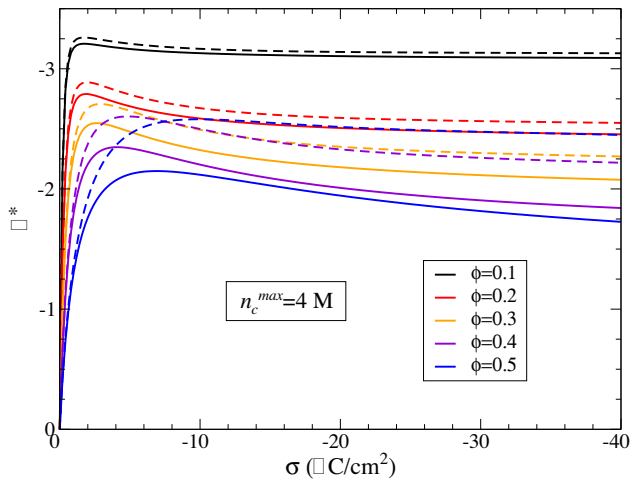}
  \caption{Scaled electrophoretic mobility against the surface charge density for different particle volume fraction values. Solid lines show the results for point-like ions. Dashed lines show the results of the FIS+L model with $n_c^{max}=4$ M.}
  \label{fgr:mob_sigmadphi_nmax4}
\end{figure}

\begin{figure}[t]
\centering
  \includegraphics[width=8.3cm]{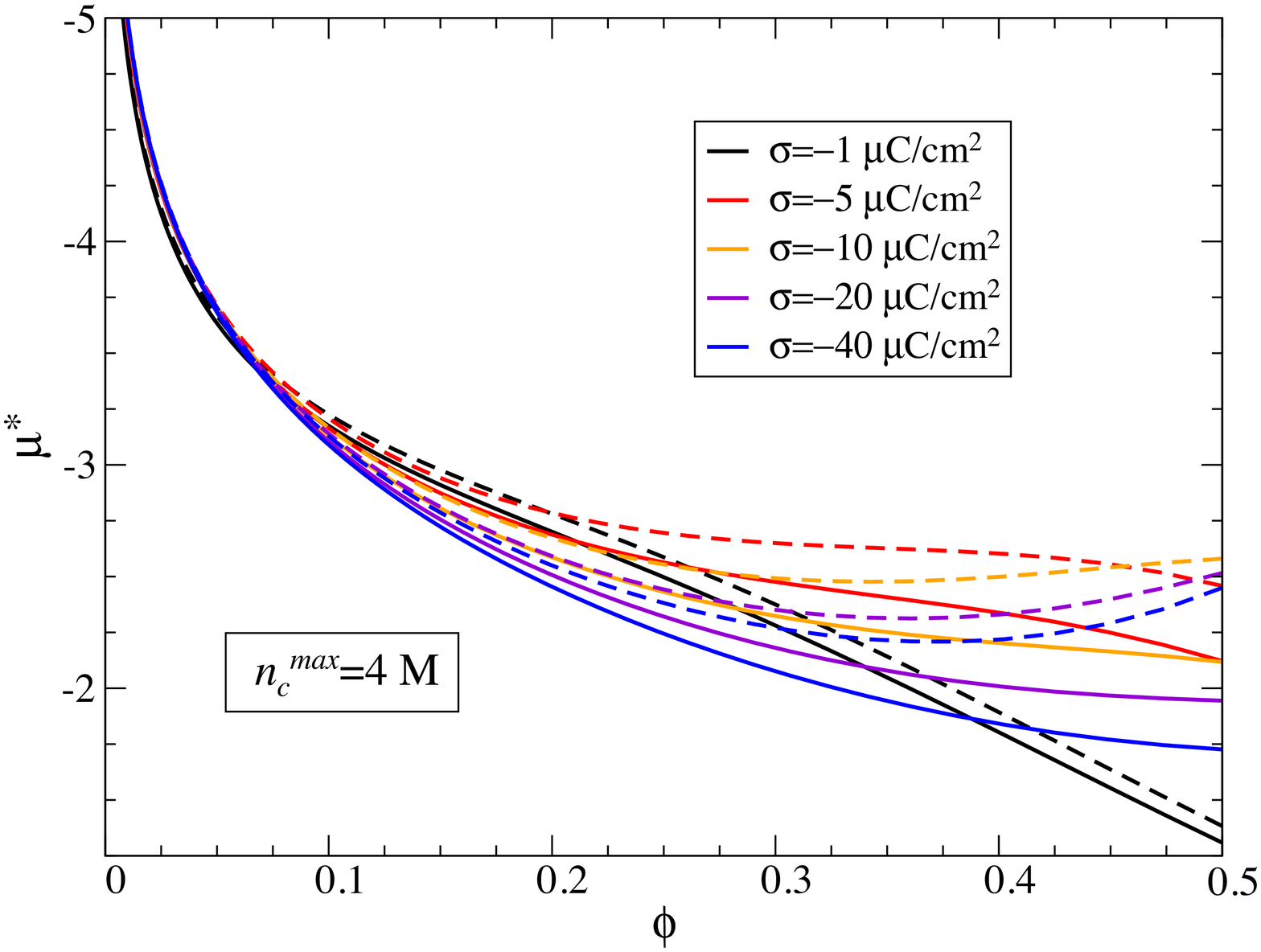}
  \caption{Scaled electrophoretic mobility against the particle volume fraction for different surface charge densities. Solid lines show the results for point-like ions. Dashed lines show the results of the FIS+L model with $n_c^{max}=4$ M.}
  \label{fgr:mob_phidsigma_nmax4}
\end{figure}

The classical behavior of the electrophoretic mobility of spherical particles in salt-free concentrated suspensions when we consider point-like counterions, PL model, is as follows: for low particle surface charges, there is a large increment of the electrophoretic mobility with the surface charge. When the particle volume fraction decreases, this increment is even larger. This behavior satisfies a H\"uckel law linearly connecting both magnitudes. When the particle charge increases, the electrophoretic mobility reaches a plateau and becomes practically independent of particle charge. This fact has been associated with the generation of a condensation layer of counterions close to the particle surface \cite{Ohshima2002}. Between these regimes there is a maximum followed by a small diminution of the electrophoretic mobility that depends on particle volume fraction. Also, the lower the particle volume fraction, the higher the electrophoretic mobility for every particle charge value. These classical behaviors are shown in solid black lines in Figs. \ref{fgr:mob_sigmaphi05} and \ref{fgr:mob_phisigma40}. The results displayed in solid colored lines in Figs. \ref{fgr:mob_sigmadphi_nmax4} and \ref{fgr:mob_phidsigma_nmax4} are also useful to understand this discussion.

If we take into account the size of the counterions, FIS model, we find deviations from the point-like case. As we can see from the results of the solid colored lines in Fig. \ref{fgr:mob_sigmaphi05}, the small diminution passed the maximum in the electrophoretic mobility tends to disappear when the ion size becomes important. Moreover, if the size of the counterions is sufficiently large, also the maximum disappears. In this case we find two different regimes for the electrophoretic mobility upon changing the surface charge of the particles: the initial large increment of the mobility with the surface charge, similar to the one for point-like ions, now followed, for higher particle charges, by another region with a small rate of increment of the electrophoretic mobility. 

When we study the behavior of the electrophoretic mobility when changing the particle volume fraction, we observe that the results of the FIS model differ from those for point-like ions, obtaining higher values for the mobility as we approach to the concentrated regime, see solid colored lines in Fig. \ref{fgr:mob_phisigma40}. Also, if the ion size is sufficiently large, we can find a broad minimum at high particle volume fraction, in contrast with the PL case.

If the counterion size approaches to zero, or equivalently $n_c^{max}\to\infty$, the results of the FIS model approximate to those of the PL model in any case. Also, for low particle charges and low particle volume fractions the results of both models are nearly the same. For the remaining situations we always observe that the FIS model predicts higher values of the electrophoretic mobility than those calculated with the classical model for point-like ions, whatever the ion size.

If we also consider the distance of closest approach of the counterions to the particle surface, FIS+L model, we obtain large differences in comparison with the PL model from moderate to high particle charges, see dashed lines in Figs. \ref{fgr:mob_sigmaphi05} and \ref{fgr:mob_sigmadphi_nmax4}. Although it is not shown in Fig. \ref{fgr:mob_sigmaphi05}, when the particle surface charge is extremely high, the electrophoretic mobility predicted with the FIS and FIS+L models must reach the same plateau value, because the distance of closest approach becomes negligible in comparison with the width of the condensation counterions layer.

When we change the particle volume fraction, dashed lines in Fig. \ref{fgr:mob_phisigma40}, the consideration of the excluded region in contact with the particle augments the effect that we observed with the FIS model. The results displayed in Fig. \ref{fgr:mob_phidsigma_nmax4} also show how for a fixed size of the counterion, the FIS+L model predicts large deviations from the PL model for concentrated suspensions (dashed lines versus solid lines). We observe that always the FIS+L model predicts equal or higher values of the electrophoretic mobility than the PL and the FIS models, whatever the ion size.

As a main conclusion from Figs. \ref{fgr:mob_sigmaphi05} to \ref{fgr:mob_phidsigma_nmax4}, we can affirm that the consideration of finite ion size effects leads to an increase of the electrophoretic mobility over the PL case, using H$^+$ ions as counterions. As the ionic concentration in the cell has been altered, we will analyze the changes in the convective fluid flow, the counterions fluxes and the perturbed counterions concentration, as well as the overall forces and polarizations induced by the electric field for the FIS and FIS+L models in comparison with the PL case.

\begin{figure}[t]
\centering
  \includegraphics[width=8.3cm]{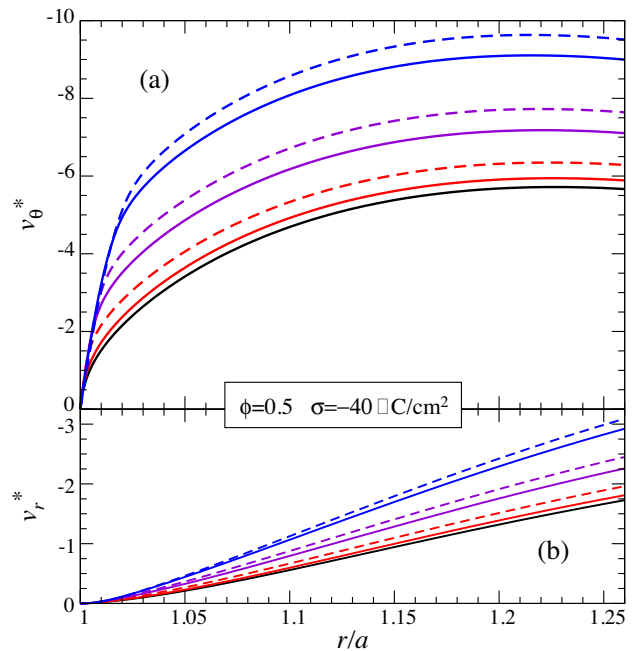}
  \caption{Scaled polar component of the fluid velocity over the particle equator, $\theta=\pi/2$, (a) and scaled radial component of the fluid velocity at the front of the particle, $\theta=\pi$, (b) along the cell for different ion sizes. Different lines have the same meaning than those shown in Fig. \ref{fgr:mob_sigmaphi05}.}
  \label{fgr:vfluid_sigma40phi05}
\end{figure}

Fig. \ref{fgr:vfluid_sigma40phi05}a shows the scaled polar component of the fluid velocity over the particle equator, $\theta=\pi/2$, and Fig. \ref{fgr:vfluid_sigma40phi05}b presents the scaled radial component of the fluid velocity at the front of the particle, $\theta=\pi$, along the cell. Solid black lines, solid colored lines and dashed lines, represent the results of the PL, FIS and FIS+L models, respectively. Different colors stand for different counterion sizes. The particle surface charge density has been chosen equal to $-$40 $\mu$C/cm$^2$, and the particle volume fraction is $\phi=0.5$, which implies a normalized cell size of $b/a=1.26$. We define the scaled fluid velocity as
\begin{equation}
\mathbf{v}^*(\mathbf{r})=\frac{3\eta e}{2\epsilon_0\epsilon_{rs}k_BTE}\mathbf{v}(\mathbf{r})
\end{equation}
where the different components of $\mathbf{v}(\mathbf{r})$ can be obtained from eqn (\ref{vOhshima}). Both quantities, $v_\theta$ at $\pi/2$ and $v_r$ at $\pi$, are of interest because they give us an idea of the magnitude of the electrophoretic velocity because they are antiparallel to it.

We can see how the polar fluid velocity increases with the distance to the particle surface, with a high rate close to the particle surface, and diminishing it as we approximate to the outer surface of the cell. In the case of the radial component, there is a linear increase after an initial slower growth rate very close to the particle surface. This behavior is the same for the PL, FIS and FIS+L models. We also observe that the numerical values obtained with the FIS+L model are larger than those obtained with the FIS model, being the latter results also larger than those obtained with the PL model, in concordance with our predictions for the mobility. 

\begin{figure}[t]
\centering
  \includegraphics[width=8.3cm]{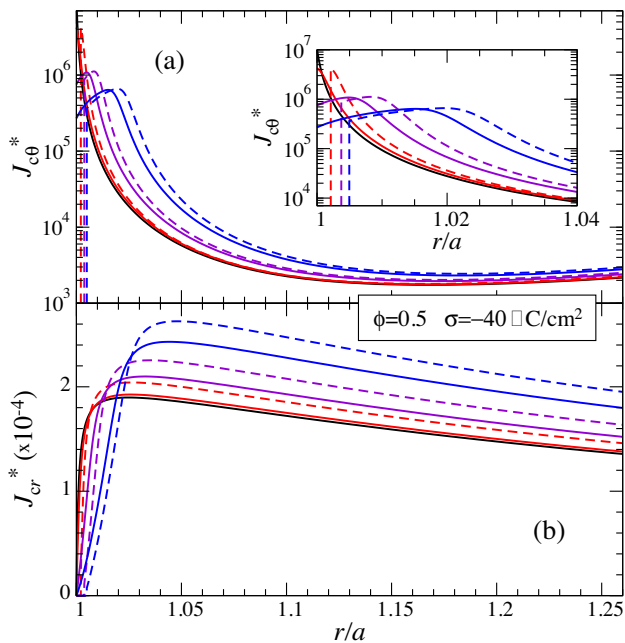}
  \caption{Scaled polar component of the flux of counterions over the particle equator, $\theta=\pi/2$, (a) and scaled radial component of the flux of counterions at the front of the particle, $\theta=\pi$, (b) along the cell for different ion sizes. Different lines have the same meaning than those shown in Fig. \ref{fgr:mob_sigmaphi05}. In the inset of Fig. \ref{fgr:flux_sigma40phi05}a we enlarge the region close to the particle surface.}
  \label{fgr:flux_sigma40phi05}
\end{figure}

Fig. \ref{fgr:flux_sigma40phi05}a shows the scaled polar component of the counterions flux over the particle equator, $\theta=\pi/2$, and Fig. \ref{fgr:flux_sigma40phi05}b presents the scaled radial component of the flux of the counterions at the front of the particle, $\theta=\pi$, along the cell. Different lines have the same meaning than in Fig. \ref{fgr:vfluid_sigma40phi05}. We define the scaled counterions flux as
\begin{equation}
\mathbf{J}_c^*(\mathbf{r})=\frac{3\eta e^3a^2}{2(\epsilon_0\epsilon_{rs}k_BT)^2E}\mathbf{J}_c(\mathbf{r})
\end{equation}
being $\mathbf{J}_c(\mathbf{r})=n_c(\mathbf{r})\mathbf{v}_c(\mathbf{r})$, where $\mathbf{v}_c(\mathbf{r})$ is obtained from eqn (\ref{nersnt-p}). We can see in Fig. \ref{fgr:flux_sigma40phi05} how the inclusion of finite ion size effects largely decreases the magnitude of the counterions fluxes close to the particle surface in both the polar and the radial directions in comparison with the PL case. In addition, the counterions fluxes are highly increased in the FIS and FIS+L cases as we move away from the particle surface because now both the counterions concentration and the counterions velocity (not shown for brevity) reach higher values.

The enhancement of the fluid velocity observed in the FIS and FIS+L cases in comparison with the PL model could be associated with the increment of the counterions fluxes in the region not in the immediate vicinity of the particle, because the ions have been expelled out due to the excluded volume effect, and consequently the electrical body force in that region is greater. However, as we can see in Table \ref{tbl:forces_pol}, the total electrical body force, $-F_e^*$, is lower in the FIS and FIS+L cases, in comparison with the PL case. This is due to a very large diminution of the electrical body force very close to the particle when the finite ion size is taken into account, not only because ions have moved away from the vicinity of the particle surface but also for the remarkable diminution of the local electric field due to the increased induced electric polarization (see Table \ref{tbl:forces_pol}). 

\begin{table}[b]
\small
  \caption{\ Scaled hydrodynamic, $F_h^*$, and electric, $F_e^*$, forces acting on the particle in the direction of the field. The total body force in the fluid is equal to $-F_e^*$ due to the electroneutrality of the cell. $\langle P_C\rangle^*$ and $\langle P_D\rangle^*$ are the scaled induced charge and dielectric polarizations in the direction of the field, respectively. All the calculations were performed at $\sigma=-40\mu$C/cm$^2$ and $\phi=0.5$.}
  \label{tbl:forces_pol}
  \begin{tabular*}{0.47\textwidth}{@{\extracolsep{\fill}}lllll}
    \hline
    Model / $n_c^{max}$ & $F_h^*$ & $F_e^*$ & $\langle P_C\rangle^*$ & $\langle P_D\rangle^*$ \\
    \hline
    PL 		  & 685 	& -696 	& 346 	& -71.7	\\
    FIS / 22 M 	  & 678 	& -677	& 350 	& -71.8	\\
    FIS / 4 M 	  & 582	& -582	& 368	& -72.6	\\
    FIS / 1.7 M 	  & 492 	& -492	& 399 	& -73.4	\\
    FIS+L / 22 M  & 524	& -524	& 365	& -73.1	\\
    FIS+L / 4 M 	  & 447	& -447	& 386	& -73.8	\\
    FIS+L / 1.7 M & 394	& -394	& 418	& -74.2	\\
    \hline
  \end{tabular*}
\end{table}

The total hydrodynamic and electric forces acting on the particle shown in Table \ref{tbl:forces_pol} can be calculated by
\begin{equation}
\mathbf{F}_h=\int_{S_p} \tilde{\sigma}^H\cdot\hat{\mathbf{r}}\ dS_p
\end{equation}
\begin{equation}
\mathbf{F}_e=\int_{S_p} \tilde{\sigma}^M\cdot\hat{\mathbf{r}}\ dS_p
\end{equation}
where $S_p$ is the surface of the solid particle, and $\tilde{\sigma}^H$ and $\tilde{\sigma}^M$ are the hydrodynamic and the Maxwell stress tensors, respectively \cite{Masliyah2006}. Evaluating these expressions, we obtain
\begin{multline}
\mathbf{F}_h=\frac{4}{3}\pi a^2E\bigg[\eta a \frac{\mathrm{d}^3h}{\mathrm{d}r^3}\bigg |_{r=a}+\eta\frac{\mathrm{d}^2h}{\mathrm{d}r^2}\bigg |_{r=a}\\ -z_cen_c^0(a)Y(a)\bigg]\hat{\mathbf{k}}
\end{multline}
\begin{equation}
\mathbf{F}_e=\frac{4}{3}\pi a^2E\sigma\frac{Y(a)}{a}\left(\frac{\epsilon_{rp}}{\epsilon_{rs}}+2\right)\hat{\mathbf{k}}
\end{equation}
where $\hat{\mathbf{k}}$ points to the direction of the macroscopic electric field. Both forces are scaled as follows
\begin{equation}
\mathbf{F}_{e,h}^*=\frac{3e}{4\pi a\epsilon_0\epsilon_{rs}k_BTE}\mathbf{F}_{e,h}
\end{equation}

The numerical values of the $F_e^*$ are negative because this force has the opposite direction of the electric field for a negative particle. The hydrodynamic force opposes the movement of the particle and therefore has the direction of the field. In the stationary state the total force acting on the particle must be zero, as we can see by summing the values of both forces in Table \ref{tbl:forces_pol}. The small numerical discrepancies observed, mainly for the PL case, can be removed by improving the mesh for the resolution of the electrokinetic equations very close to the particle surface although a large computational time is required. 

It is worthwhile to mention that both the hydrodynamic and the electric forces calculated in Table \ref{tbl:forces_pol} for the FIS and FIS+L models are lower than those of the PL case. The electric force acting on the particle has a driving and a relaxation contribution. As the driving force is constant in the study displayed in Table \ref{tbl:forces_pol}, the diminution of the total electric force is related with a change of the relaxation force as finite ion size is taken into account.

This relaxation force will depend on the electric dipole moment induced on the particle and its double layer by the electric field. A related quantity is the average induced dipole moment density whose components are: the charge polarization, $\langle \mathbf{P}_C\rangle$, and the induced dipole moment density arising from the polarization of the dielectric continuum of the medium and the particles, $\langle \mathbf{P}_D\rangle$
\begin{equation}\label{pol_charge}
\langle \mathbf{P}_C\rangle=\left\langle \frac{1}{V_{cell}}\int_{V_{cell}}\mathbf{r}z_ce\delta n_c(\mathbf{r})\mathrm{d}V \right\rangle
\end{equation}
\begin{equation}
\langle \mathbf{P}_D\rangle=\left\langle \frac{-1}{V_{cell}}\int_{V_{cell}}\mathbf{r}(\epsilon(\mathbf{r})-\epsilon_0)\nabla\delta\Psi(\mathbf{r})\mathrm{d}V \right\rangle
\end{equation}
where
\begin{equation}
\epsilon(\mathbf{r})=
\begin{cases}
\epsilon_{rp}\epsilon_0 & \text{ \ } \mathbf{r}\in V_p \\
& \\
\epsilon_{rs}\epsilon_0 & \text{ \ } \mathbf{r}\in V_s
\end{cases}
\end{equation}
and $V_p$ and $V_s$ are the particle and solution volumes in the cell, respectively. According to the procedure developed for AC electric fields and point-like ions by Bradshaw-Hajek \textit{et al.} \cite{Bradshaw2010}, particularized to its DC limit and also accounting for the distance of closest approach of the counterions to the particle surface, the latter equations become
\begin{multline}
\langle \mathbf{P}_C\rangle=-\epsilon_0\epsilon_{rs}\bigg[1-\frac{\mathrm{d}Y}{\mathrm{d}r}\bigg |_{r=b}+\left(\frac{a+R}{b}\right)^3 \\
\times\left(\frac{\mathrm{d}Y}{\mathrm{d}r}\bigg |_{r=a+R^+}-\frac{Y(a+R^-)}{a+R}\right)\bigg] \langle \mathbf{E}\rangle
\end{multline}
\begin{equation}
\langle \mathbf{P}_D\rangle=\epsilon_0\left[(\epsilon_{rp}-\epsilon_{rs})\phi\frac{Y(a)}{a} +(\epsilon_{rs}-1)\right]\langle \mathbf{E}\rangle
\end{equation}

We show in Table \ref{tbl:forces_pol} the scaled average polarization contributions, $\langle \mathbf{P}_C\rangle^*$ and $\langle \mathbf{P}_D\rangle^*$, calculated by
\begin{equation}
\langle \mathbf{P}_{C,D}\rangle^*=\frac{\langle \mathbf{P}_{C,D}\rangle}{\epsilon_0E}
\end{equation}

The charge polarization takes positive values and therefore the induced dipole moment generates an electric field which opposes the external one, thus penalizing the particle movement. On the contrary, the induced dipole moment due to the dielectric polarization has the opposite direction, reinforcing the effect of the external electric field on the particle movement. As the charge polarization contribution is considerably larger than the dielectric polarization one, see Table \ref{tbl:forces_pol}, the effect on the particle movement will be a net relaxation force that opposes the electric driving force, yielding a smaller total electric force in the FIS and FIS+L models, in comparison with the PL case (see Table \ref{tbl:forces_pol}).

\begin{figure}[t]
\centering
  \includegraphics[width=8.3cm]{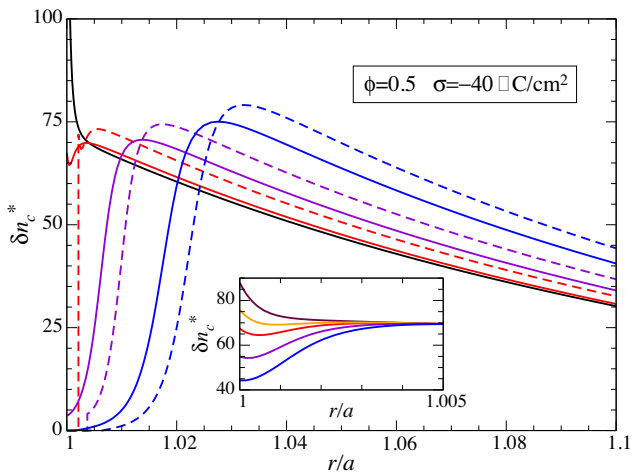}
  \caption{Scaled perturbed counterion concentration in the direction of the field ($\theta=0$) along the cell for different ion sizes. Different lines have the same meaning than those in Fig. \ref{fgr:mob_sigmaphi05}. In the inset, $\delta n_c^*$ versus $r/a$ in the neighborhood of the particle surface with the FIS model. Colored lines stand for $n_c^{max}=15$, 18, 22, 25 and 30 M, respectively, from bottom to top.}
  \label{fgr:deltanc_sigma40phi05}
\end{figure}

In Fig. \ref{fgr:deltanc_sigma40phi05} we observe the scaled perturbed counterion concentration in the direction of the field ($\theta=0$) along the cell. This quantity is responsible of the charge polarization contribution as we can see in eqn (\ref{pol_charge}). Solid black lines, solid colored lines and dashed lines, represent the results of the PL, FIS and FIS+L models, respectively. Different colors stand for different counterion sizes. The perturbed counterion concentration is scaled as
\begin{equation}
\delta n_c^*(r)=\frac{ea}{\epsilon_0\epsilon_{rs}E}\delta n_c(r)
\end{equation}
where, using eqns (\ref{echempot}), (\ref{deltamuc}) and (\ref{deltapsi}), we have  
\begin{equation}
\delta n_c(r)=\frac{z_ce}{k_BT} n_c^0(r)\left(Y(r)-\phi_c(r)\right) E\cos\theta
\end{equation}

In the inset of Fig. \ref{fgr:deltanc_sigma40phi05} we represent $\delta n_c^*$ versus $r/a$ in the neighborhood of the particle with the FIS model for $n_c^{max}=15$, 18, 22, 25 and 30 M, respectively, from bottom to top, just to clarify the transition between the PL and FIS cases.

We observe that the perturbed counterion concentration decreases asymptotically to zero at the outer surface of the cell ($b/a=1.26$) in all cases according to Shilov-Zharkikh-Borkovskaya boundary condition, eqn (\ref{shilovzarbor}). In the PL case, we obtain an excess of counterions at the rear of the particle, $\theta=0$, and a defect of counterions at the front of the particle, $\theta=\pi$, due to a counterions migration from the front to the rear of the particle when the external electric field is applied. This excess of counterions is mainly located very close to the particle surface and generates an electric dipole moment that points to the direction of the electric field. 

When we take into account the finite counterions size, FIS model, the excess of counterions is lower near the particle surface in comparison with the PL case, see Fig. \ref{fgr:deltanc_sigma40phi05}. As we noted in a previous work (Fig. 2 of ref. \cite{Roa2011}), the equilibrium concentration profile shows a counterions condensate that increases its width upon increasing the size of the counterions because of steric reasons. Therefore, when the external electric field is applied, the excess of counterions cannot be located in the condensate, full of counterions mainly for high particle charges. This is the reason why the main region of excess of counterions is now found at farther distances from the particle surface. When we consider the excluded region in contact with the particle, FIS+L model, the perturbed concentration profile is shifted to larger distances from the particle surfaces as we see in Fig. \ref{fgr:deltanc_sigma40phi05}. In all the cases studied, the charge contribution to the induced electric dipole moment points to the direction of the electric field which penalizes the movement of the particle by the relaxation effect \cite{Masliyah2006}. 

According to Table \ref{tbl:forces_pol}, the charge polarization contribution is larger the larger the ion size, and even more if we take into account the distance of closest approach of the counterions to the particle surface. This fact is in concordance with an increase in the charge contribution to the induced electric dipole moment that we can associate to the charge redistribution depicted in Fig. \ref{fgr:deltanc_sigma40phi05} due to the excluded volume effect. 

Summarizing, we have seen that when we introduce the ion size effects there is a remarkable diminution of the total electric force acting on the particle (see Table \ref{tbl:forces_pol}). This decrease grows with the ion size and with the inclusion of the excluded region. The diminution of the total electric force must be accompanied, in the stationary state, by a corresponding diminution of the total hydrodynamic force. According to the theory of classical electrokinetics \cite{Masliyah2006}, the total hydrodynamic force could be decomposed in a viscous drag and a electrophoretic retardation contribution. In this frame, the above mentioned decrease of the electric body force as ions size effects are considered provokes a diminution of the electrophoretic retardation force, which opposes to the movement of the particle, as it happens in the FIS and FIS+L cases. However, a complete explanation of the observed increase in the electrophoretic mobility will force us to study the transient regime after the application of the external field and the evolution of the different forces involved until the stationary state is reached. In the case we are concerned in this work, the final result is that the electrophoretic mobility of the particle is higher in the FIS and the FIS+L cases in comparison with the PL model using H$^+$ ions as counterions, being these ions commonly found in many experimental salt-free suspensions. For other ionic species the behaviors observed can be different, depending on the diffusion coefficient of the ion, because the various contributions to the total force will be altered. The influence of the mobility of the ions will be addressed in a future work in connection with the corresponding experimental results.

\subsection{Electrical conductivity} 

\begin{figure}[t]
\centering
  \includegraphics[width=8.3cm]{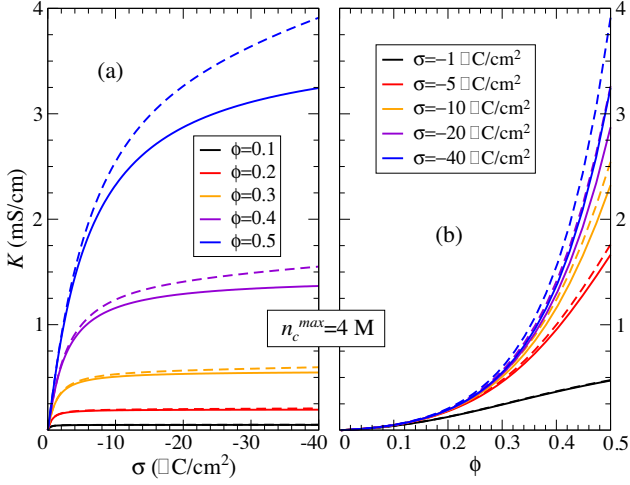}
  \caption{(a) Electrical conductivity against the surface charge density for different particle volume fraction values. (b) Electrical conductivity against the particle volume fraction for different surface charge densities. Solid lines show the results for point-like ions. Dashed lines show the results of the FIS+L model.}
  \label{fgr:cond_sigmaphi_nmax4}
\end{figure}

\begin{figure}[t]
\centering
  \includegraphics[width=8.3cm]{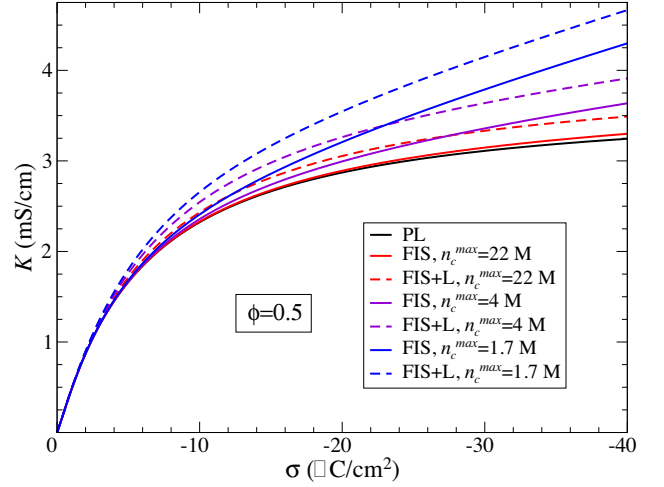}
  \caption{Electrical conductivity against the particle surface charge density for different ion sizes, considering (dashed lines) or not (solid lines) the excluded region in contact with the particle. Black lines show the results for point-like ions.}
  \label{fgr:cond_sigmaphi05}
\end{figure}

\begin{figure}[t]
\centering
  \includegraphics[width=8.3cm]{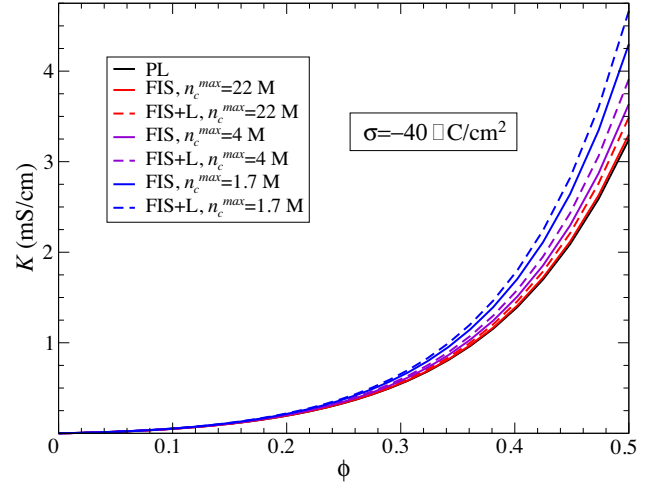}
  \caption{Electrical conductivity against the particle volume fraction for different ion sizes, considering (dashed lines) or not (solid lines) the excluded region in contact with the particle. Black lines show the results for point-like ions.}
  \label{fgr:cond_phisigma40}
\end{figure}

The electrical conductivity of salt-free concentrated suspensions with point-like ions behaves classically as follows: the conductivity increases as surface charge density increases for any volume fraction \cite{Carrique2006}. For large particle charges, the conductivity tends to reach a plateau because of the classic counterion condensation effect \cite{Ohshima2002}: it appears that a critical particle charge density exists beyond which there is no appreciable influence on the conductivity. Once this critical charge value is attained, increasing the amount of counterions by raising the surface charge even more simply feeds the condensation region, where a high accumulation of counterions takes place close to the particle surface, leaving the charge and potential outside that region virtually unchanged. There is also a conductivity enhancement with particle volume fraction because the increasing number of the double-layer mobile ions is not offset by the presence of the nonconducting volume occupied by the particles in the unit volume. These behaviors are shown in solid colored lines in Fig. \ref{fgr:cond_sigmaphi_nmax4} and in solid black lines in Figs. \ref{fgr:cond_sigmaphi05} and \ref{fgr:cond_phisigma40}.

When we take into account the finite size of the counterions (FIS model, solid colored lines in Figs. \ref{fgr:cond_sigmaphi05} and \ref{fgr:cond_phisigma40}) and the distance of closest approach of the counterions to the particle surface (FIS+L model, dashed lines in Figs. \ref{fgr:cond_sigmaphi_nmax4}, \ref{fgr:cond_sigmaphi05} and \ref{fgr:cond_phisigma40}) we observe similar behaviors but the numerical values of the electrical conductivity are always higher for any counterion size for moderate to high particle charges in concentrated suspensions. If the counterion size approaches to zero, or equivalently $n_c^{max}\to\infty$, the results of the FIS and FIS+L models approximate to those of the PL model. Also, for low particle charges and low particle volume fractions the results of the three models are almost coincident. 

According to Fig. \ref{fgr:flux_sigma40phi05} the counterions fluxes not in the immediate vicinity of the particle have been enhanced in the FIS and FIS+L model in comparison with the PL case. This enhancement gives rise to a larger conductivity of the suspension, because it predominates over the larger PL counterions fluxes very close to the particle surface. As these augmented counterions fluxes in the FIS and FIS+L models grow with the counterion size and with the inclusion of the excluded region, the electric conductivity increases as well (see Figs. \ref{fgr:cond_sigmaphi_nmax4}, \ref{fgr:cond_sigmaphi05} and \ref{fgr:cond_phisigma40}). 

Although all the conductivity calculations have been performed with H$^+$ counterions, we have checked that the conductivity behavior shown before maintains for other counterions species with different diffusion coefficients.

\section{Conclusions} \label{sec:conclusions}

In this work we have studied the influence of finite ion size corrections on the electrophoretic mobility of spherical particles and the electrical conductivity in salt-free concentrated suspensions with H$^+$ added counterions, although the model is also valid for different ionic species. The resulting model is based on a mean-field approach that has reasonably succeeded in modeling electrokinetic and rheological properties of concentrated suspensions. We have used a cell model approach to account for particle-particle interactions, and derived a DC electrokinectic model which include such ion size effects. The theoretical procedure has followed that by Carrique \textit{et al.} \cite{Carrique2006} with the additional inclusion of finite size counterions \cite{Borukhov1997} and an excluded region of closest approach of the ions to the particle surface \cite{ArandaRascon2009b}. 

The results have shown that the finite ion size effect has to be taken into account for moderate to high particle charges in concentrated suspensions, and even more if a distance of closest approach of the ions to the particle surface is considered. In the common case of H$^+$ counterions, we have found a larger increase of the electrophoretic mobility and the suspension conductivity the larger the ion size. The effect is more noticeable if we take into account an excluded region free of ions in contact with the particle.

The DC electrokinetic model presented in this paper will be used to develop a theoretical model on the response of a salt-free concentrated suspension to an AC electric field including ion size effects. Experimental results concerning the DC electrophoretic mobility, dynamic electrophoretic mobility, electrical conductivity and dielectric response with different counterions species should be compared with the predictions of the latter models to test them. To perform such comparisons, concentrated suspensions of highly charged particles are required, which classically have been very difficult to synthesize. This is probably the reason that explains the lack of experimental studies that could be used to test our predictions. Nowadays, existing highly charged sulfonated polystyrene latexes could be good candidates. As the authors shown in a previous paper \cite{Roa2011b}, realistic salt-free suspensions which include water dissociation ions and those generated by atmospheric carbon dioxide contamination, in addition to the added counterions released by the particles to the solution, should be considered to get closer to the experimental results. The inclusion of these chemical reactions in an electrokinetic model is not trivial and require a proper electrokinetic model that extends the present one. These theoretical and experimental tasks will be addressed by the authors in the near future. 

\section*{Acknowledgements}
Junta de Andaluc\'ia, Spain (Project P08-FQM-3779), and MICINN, Spain (Project FIS2010-18972), co-financed with FEDER funds by the EU.


\begin{thebibliography}{99}

\bibitem{Dukhin1974} S. S. Dukhin and V. N. Shilov, \textit{Dielectric phenomena and the double layer in disperse systems and polyelectrolytes}, Wiley, New York, 1974.

\bibitem{Lyklema1995} J. Lyklema, \textit{Fundamentals of interface and colloid science: vol. II, Solid-liquid interfaces}, Academic Press, London, 1995.

\bibitem{Masliyah2006} J. H. Masliyah and S. Bhattacharjee, \textit{Electrokinetic and colloid transport phenomena}, Wiley-Interscience, New Jersey, 2006.

\bibitem{OBrien1978} R. W. O'Brien and L. R. White, \textit{J. Chem. Soc., Faraday Trans. 2}, 1978, \textbf{74}, 1607-1626.

\bibitem{OBrien1981} R. W. O'Brien, \textit{J. Colloid Interface Sci.}, 1981, \textbf{81}, 234-248.

\bibitem{OBrien1990} R. W. OÕBrien, \textit{J. Fluid Mech.}, 1990, \textbf{212}, 81-93.

\bibitem{Dukhin1999} A. S. Dukhin, H. Ohshima, V. N. Shilov, and P. J. Goetz, \textit{Langmuir}, 1999, \textbf{15}, 3445-3451.

\bibitem{Ohshima1997} H. Ohshima, \textit{J. Colloid Interface Sci.}, 1997, \textbf{188}, 481-485.

\bibitem{Ohshima1999} H. Ohshima, \textit{J. Colloid Interface Sci.}, 1999, \textbf{212}, 443-448.

\bibitem{Carrique2003b} F. Carrique, F. J. Arroyo, M. L. Jim\'enez and A. V. Delgado, \textit{J. Chem. Phys.}, 2003, \textbf{118}, 1945-1956.

\bibitem{Delgado2005} A. V. Delgado, S. Ahualli, F. J. Arroyo and F. Carrique, \textit{Colloids Surf. A}, 2005, \textbf{267}, 95-102.

\bibitem{Cuquejo2006} J. Cuquejo, M. L. Jim\'enez, A. V. Delgado, F. J. Arroyo and F. Carrique, \textit{J. Phys. Chem. B}, 2006, \textbf{110}, 6179-6189.

\bibitem{Reiber2007} H. Reiber, T. K\"oller, E. Ruiz-Reina, T. Palberg, R. Piazza and F. Carrique, \textit{J. Colloid Interface Sci.}, 2007, \textbf{309}, 315-322.

\bibitem{Carrique2003} F. Carrique, F. J. Arroyo, M. L. Jim\'enez and A. V. Delgado, \textit{J. Phys. Chem. B}, 2003, \textbf{107}, 3199-3206.

\bibitem{Sood1991} A. K. Sood, in \textit{Solid State Physics}, ed. H. Ehrenreich and D. Turnbull, Academic Press, 1991, vol. 45, p. 1.

\bibitem{Medebach2003} M. Medebach and T. Palberg, \textit{J. Chem. Phys.}, 2003, \textbf{119}, 3360-3370.

\bibitem{Palberg2004} T. Palberg, M. Medebach, N. Garbow,  M. Evers, A. B. Fontecha, H. Reiber and E. Bartsch, \textit{J. Phys.: Condens. Matter}, 2004, \textbf{16}, S4039-S4050.

\bibitem{Medebach2004} M. Medebach and T. Palberg, \textit{J. Phys.: Condens. Matter}, 2004, \textbf{16}, 5653-5658.

\bibitem{Ohshima2002} H. Ohshima, \textit{J. Colloid Interface Sci.}, 2002, \textbf{248}, 499-503.

\bibitem{Chiang2006} C. P. Chiang, E. Lee, Y. Y. He and J. P. Hsu, \textit{J. Phys. Chem. B}, 2006, \textbf{110}, 1490-1498.

\bibitem{Carrique2006} F. Carrique, E. Ruiz-Reina, F. J. Arroyo and A. V. Delgado, \textit{J. Phys. Chem. B}, 2006, \textbf{110}, 18313-18323.

\bibitem{LozadaCassou1999} M. Lozada-Cassou, E. Gonz\'alez-Tovar and W. Olivares, \textit{Phys. Rev. E}, 1999, \textbf{60}, R17-20.

\bibitem{Lobaskin2007} V. Lobaskin, B. D\"unweg, M. Medebach, T. Palberg and C. Holm, \textit{Phys. Rev. Lett.}, 2007, \textbf{98}, 176105.

\bibitem{Chatterji2007} A. Chatterji and J. Horbach, \textit{J. Chem. Phys.}, 2007, \textbf{126}, 064907.

\bibitem{MartinMolina2009} A. Mart\'in-Molina, C. Rodr\'iguez-Beas, R. Hidalgo-\'Alvarez and M. Quesada-P\'erez \textit{J. Phys. Chem. B}, 2009, \textbf{113}, 6834-6839.

\bibitem{Pagonabarraga2010} I. Pagonabarraga, B. Rotenberg and D. Frenkel \textit{Phys. Chem. Chem. Phys.}, 2010, \textbf{12}, 9566-9580.

\bibitem{Lyklema2009} J. Lyklema, \textit{Adv. Colloid Interface Sci.}, 2009, \textbf{147}, 205-213.

\bibitem{Bikerman1942} J. J. Bikerman, \textit{Philos. Mag.}, 1942, \textbf{33}, 384-397.

\bibitem{KraljIglic1996} V. Kralj-Iglic and A. Iglic, \textit{J. Phys. II France}, 1996, \textbf{6}, 477-491.

\bibitem{Borukhov1997} I. Borukhov, D. Andelman and H. Orland, \textit{Phys. Rev. Lett.}, 1997, \textbf{79}, 435-438.

\bibitem{Borukhov2004} I. Borukhov, \textit{J. Polym. Sci. B Polym. Phys.}, 2004, \textbf{42}, 3598-3615.

\bibitem{Kilic2007} M. S. Kilic, M. Z. Bazant and A. Ajdari, \textit{Phys. Rev. E}, 2007, \textbf{75}, 021503.

\bibitem{ArandaRascon2009} M. J. Aranda-Rasc\'on, C. Grosse, J. J. L\'opez-Garc\'ia and J. Horno, \textit{J. Colloid Interface Sci.}, 2009, \textbf{335}, 250-256.

\bibitem{LopezGarcia2010} J. J. L\'opez-Garc\'ia, M. J. Aranda-Rasc\'on, C. Grosse, and J. Horno, \textit{J. Phys. Chem. B}, 2010, \textbf{114}, 7548-7556.

\bibitem{Bazant2009} M. Z. Bazant, M. S. Kilic, B. D. Storey and A. Ajdari, \textit{Adv. Colloid Interface Sci.}, 2009, \textbf{152}, 48-88.

\bibitem{IbarraArmenta2009} J. Ibarra-Armenta, A. Mart\'in-Molina and M. Quesada-P\'erez, \textit{Phys. Chem. Chem. Phys.}, 2009, \textbf{11}, 309-316.

\bibitem{ArandaRascon2009b} M. J. Aranda-Rasc\'on, C. Grosse, J. J. L\'opez-Garc\'ia and J. Horno, \textit{J. Colloid Interface Sci.}, 2009, \textbf{336}, 857-864.

\bibitem{Khair2009} A. S. Khair and T. M. Squires, \textit{J. Fluid Mech.}, 2009, \textbf{640}, 343Ð356.

\bibitem{LopezGarcia2011} J. J. L\'opez-Garc\'ia, M. J. Aranda-Rasc\'on, C. Grosse, and J. Horno, \textit{J. Colloid Interface Sci.}, 2011, \textbf{356}, 325-330.

\bibitem{Roa2011} R. Roa, F. Carrique and E. Ruiz-Reina, \textit{Phys. Chem. Chem. Phys.}, 2011,  \textbf{13}, 3960-3968.

\bibitem{Zholkovskij2007} E. K. Zholkovskij, J. H. Masliyah, V. N. Shilov and S. Bhattacharjee, \textit{Adv. Colloid Interface Sci.}, 2007, \textbf{134-135}, 279-321.

\bibitem{Happel1958} J. Happel, \textit{AIChE J.}, 1958, \textbf{4}, 197-201.

\bibitem{Shilov1981} V. N. Shilov, N. I.  Zharkikh and Yu. B. Borkovskaya, \textit{Colloid J.}, 1981, \textbf{43}, 434-438.

\bibitem{Kuwabara1959} S. Kuwabara, \textit{J. Phys. Soc. Jpn.}, 1959, \textbf{14}, 527-532.

\bibitem{Ahualli2009} S. Ahualli, M. L. Jim\'enez, F. Carrique and A. V. Delgado, \textit{Langmuir}, 2009,  \textbf{25}, 1986-1997.

\bibitem{Carrique2008} F. Carrique, E. Ruiz-Reina, F. J. Arroyo, M. L. Jim\'enez and A. V. Delgado, \textit{Langmuir}, 2008, \textbf{24}, 2395-2406.

\bibitem{Carrique2005} F. Carrique, J. Cuquejo, F. J. Arroyo, M. L. Jim\'enez and A. V. Delgado, \textit{Adv. Colloid Interface Sci.}, 2005, \textbf{118}, 43-50.

\bibitem{Kierzenka2001} J. Kierzenka and L. F. Shampine, \textit{ACM Trans. Math. Softw.}, 2001, \textbf{27}, 299-316.

\bibitem{Israelachvili1992} J. N. Israelachvili, \textit{Intermolecular and surface forces}, Academic Press, London, 1992.

\bibitem{Bradshaw2010} B. H. Bradshaw-Hajek, S. J. Miklavcic and L. R. White, \textit{Langmuir}, 2010,  \textbf{26}, 7875-7884.

\bibitem{Roa2011b} R. Roa, F. Carrique and E. Ruiz-Reina, \textit{Phys. Chem. Chem. Phys.}, 2011,  \textbf{13}, 9644-9654.

\end{thebibliography}
\end{document}